\documentclass[11pt]{article}
\usepackage{latexsym}
\usepackage{amssymb}
\usepackage{epsf}
\usepackage{amsmath}
\usepackage{slashed}
\usepackage[hypertex]{hyperref}
\usepackage{graphicx}% Include figure files
\renewcommand{\thefootnote}{\fnsymbol{footnote}}

\begin{document}

\begin{titlepage}

\begin{center}

\hfill UT-11-24 \\
\hfill July, 2011 \\

\vspace{3cm}

{\bf\Large
Singlet Boson in Supersymmetric Model \\
as a Mimic of the Standard Model Higgs at the LHC}
\\

\vspace*{1.5cm}
{\large 
Masaki Asano, Takeo Moroi and Norimi Yokozaki
} \\
\vspace*{0.5cm}

{\it Department of Physics, University of Tokyo, Tokyo 113-0033, Japan}
\vspace*{0.5cm}

\end{center}

\vspace*{1.0cm}

\begin{abstract}
  We show that a gauge singlet scalar boson in low-energy
  supersymmetric model may behave as the standard model (SM) Higgs
  boson if the singlet couples to (heavy) vector-like colored
  particles.  In this case, the SM-Higgs-like signal at the LHC can be
  mimicked by the singlet production process for wide range of the
  singlet mass.
\end{abstract} 

\end{titlepage}

\baselineskip=18pt
\setcounter{page}{1}
\pagestyle{plain}

\renewcommand{\thefootnote}{\#\arabic{footnote}}
\setcounter{footnote}{0}

In the field of particle physics, Higgs boson is the most important
particle which should be experimentally found.  Discovery of the Higgs
boson itself is of course a great progress because it confirms the
Higgs mechanism as the origin of electroweak symmetry breaking.  Once
discovered, in addition, we will also be able to acquire information
about the physics beyond the standard model (SM) by detailed study of
the Higgs boson.  This is because the properties of Higgs, in
particular the Higgs mass, are strongly dependent on how the SM is 
embedded into the new physics model.

The most prominent example of physics beyond the SM is the low-energy supersymmetry (SUSY); the
minimal SUSY SM (MSSM) predicts the existence of ``light'' Higgs
boson.  At the tree level, the MSSM predicts the Higgs mass to be
smaller than the $Z$-boson mass.  Even if loop corrections are taken
into account, the upper bound on the Higgs mass is $\sim 135\ {\rm
  GeV}$ assuming that the stop mass is lighter than $\sim 2\ {\rm
  TeV}$ \cite{Carena:2002es}.  In many extended SUSY models, there still
exists light Higgs boson.  For example, in the next-to-the MSSM
(NMSSM), which is the model with gauge-singlet superfield, the
lightest Higgs mass is predicted to be smaller than $140\ {\rm GeV}$
as far as the perturbativity of coupling constants up to the grand unified theory 
(GUT) scale is assumed \cite{Ellwanger:2006rm}.  Thus, the existence of the light Higgs 
boson is a crucial check point of low-energy SUSY.

The LHC experiment is about to discover the Higgs boson if its mass is
within the accessible range.  Even if a Higgs-like object is found,
however, it would be still a question whether it is really the Higgs
boson which is responsible for the electroweak symmetry breaking.  In
the study of low-energy SUSY, this is an important question because
some class of SUSY models may be discriminated based on the properties
of the observe Higgs-like object.  Thus we should clarify if there is
any possibility of a particle which may mimic SM-Higgs-like
signals at the LHC.

In non-SUSY case, there are such possibilities \cite{Higgs_like,singlet_Higgs_like}. 
In particular, the Higgs-like signal may show up at the LHC if there
exists a singlet boson $s$ which couples to vector-like colored
fermions (or directly couples to gluon via the dimension 5 interaction
of $sG_{\mu\nu}G_{\mu\nu}$, with $G_{\mu\nu}$ being the field strength
of the gluon field) \cite{singlet_Higgs_like}.  Because a singlet chiral multiplet
appears in various SUSY models, including the NMSSM, we pursue
this possibility in the framework of SUSY model .

In this letter, we will show that the SM-Higgs-like signal at the LHC
can be easily mimicked if there exists a singlet field which couples
to a new vector-like colored chiral supermultiplets.  Thus, even if
the LHC experiment finds a SM-Higgs-like signal at the mass region
above the SUSY Higgs mass bound, low-energy SUSY is still a
possibility.  For this mechanism to work, we also show that there
should exist colored particles with their masses smaller than
$500-600\ {\rm GeV}$.

First let us briefly summarize the basic idea how the SM-Higgs-like
signal is mimicked in the SUSY model with singlet.  As is well-known,
the most efficient process of producing Higgs boson at the LHC is the gluon
fusion process, which is induced by the top-loop diagram.  Thus, if
the singlet field couples to colored particles, then we expect a
significant production cross section for the singlet field due to the
gluon fusion.  In addition, in SUSY models with singlet (as in the
case of NMSSM), the singlet mixes with the SM-like Higgs
boson due to the superpotential and SUSY breaking interactions. Consequently,
if the singlet dominantly decays via the mixing, then the LHC
signal of the singlet production is like that of the SM-Higgs
production.

To see this can really happen in SUSY model, we adopt the simplest set
up.  We introduce the gauge-singlet superfield $S$ as well as the
following vector-like chiral multiplets: $D_i$ $({\bf 3},{\bf
  1},-\frac{1}{3})$, $\bar{D}_i$ $(\bar{\bf 3},{\bf 1},\frac{1}{3})$,
$L_i$ $({\bf 1},{\bf 2},-\frac{1}{2})$, and $\bar{L}_i$ $({\bf 1},{\bf
  2},\frac{1}{2})$, where the representations for the SM gauge group
$SU(3)_C\times SU(2)_L\times U(1)_Y$ are shown in the parenthesis.
Because GUT is one of the strong motivation to
consider SUSY, we have introduced $L$ and $\bar{L}$ so that the newly
introduced fields can be embedded into the complete multiplet of
$SU(5)_{\rm GUT}$.  As we will show below, $L$ and $\bar{L}$ are not
important for the enhancement of the singlet-boson production at the
LHC.  In our analysis, we introduce $N_{\bf 5}$-pairs of vector-like
multiplets; the index $i$ runs from $1$ to $N_{\bf 5}$.  Notice that
the perturbativity of the gauge coupling constant breaks down at
high-energy scale if $N_{\bf 5}$ is too large.  Requiring the
perturbativity up to the GUT scale, we require $N_{\bf 5}\leq 4$ in our
analysis.

Motivated by the NMSSM, we adopt the following form of the
superpotential for the singlet field:
\begin{eqnarray}
  W &\supset& 
  \lambda S H_u H_d + \frac{1}{3} \kappa S^3
  + y_D S \bar{D}_i D_i
  + y_L S \bar{L}_i L_i,
  \label{eq:W_NMSSM}
\end{eqnarray}
where $H_d$ and $H_u$ are down- and up-type Higgses, respectively.
For simplicity, the Yukawa coupling constants for the $S\bar{D}_iD_i$
and $S\bar{L}_iL_i$ terms are taken to be $i$-independent; this may be due to
$SU(N_{\bf 5})$ flavor symmetry.  We have checked that our conclusion
does not significantly change even if we do not adopt this assumption.
In addition, the relevant part of the soft-SUSY breaking term is given
by
\begin{eqnarray}
  {\cal L}_{\rm soft} &\supset& 
  - m_{H_d}^2 |H_d|^2 - m_{H_u}^2 |H_u|^2 - m_{S}^2 |S|^2 
  - \left( \lambda A_\lambda S H_u H_d
    + \frac{1}{3} \kappa A_\kappa S^3 + \rm{h.c.}\right).
\end{eqnarray}
In our analysis, for simplicity, we fix some of the parameters as
\begin{eqnarray}
  A_\lambda = 500\ {\rm GeV}, \quad A_\kappa = -100\ {\rm GeV}.
\end{eqnarray}
The Higgs potential is also sensitive to some of the
parameters in the MSSM Lagrangian.  Here, we take one of the important
parameter, the tri-linear coupling for the stop (normalized by the top
Yukawa coupling constant), as $A_t=2\ {\rm TeV}$, while other
tri-linear scalar coupling parameters in the MSSM sector are taken to be $0$. 
Here, we have adopted a relatively large value of the $A_t$-parameter in order
to satisfy the LEP bound for the lightest Higgs boson
mass\cite{Barate:2003sz}.  Furthermore, the masses of all the MSSM
sfermions are taken to be $1\ {\rm TeV}$.\footnote
{We assume that the gauginos are so heavy that they do not affect the
  following analysys.}

Then, with other parameters being properly chosen, Higgs bosons acquire
vacuum expectation values; rest of the parameters (i.e., $\lambda$,
$\kappa$, $m_{H_d}^2$, $m_{H_u}^2$, $m_{S}^2$) are determined 
by fixing the following quantities:
\begin{eqnarray*}
  m_Z, \quad 
  \tan\beta =\langle H_u\rangle /  \langle H_d\rangle, \quad 
  v_s = \langle S\rangle, \quad 
  \mu_{\rm eff} = \lambda v_s, \quad 
  m_{h_2},
\end{eqnarray*}
where $\langle\cdots\rangle$ denotes vacuum expectation value.  Here,
$\mu_{\rm eff}$ plays the role of SUSY invariant Higgs mass (so-called
$\mu$-parameter) in the MSSM.  In addition, once the singlet field
gets a vacuum expectation value, then the newly introduced vector
multiplets also become massive.\footnote
{One may also introduce a bare mass term for the vector-like
  multiplets.  Extension of our analysis to such a case is
  straightforward.}

In the present model, the Higgs bosons are composed of five neutral
and two charged scalar bosons: three CP-even Higgs bosons $h_a$ (with
$m_{h_1}<m_{h_2}<m_{h_3}$), two CP-odd Higgs bosons and charged Higgs
bosons.\footnote
{We assume that the CP violation in the Higgs sector is negligible so
  that there is no mixing between the CP-even and CP-odd sectors.}
Because we are interested in the SM-Higgs-like signal at the LHC, 
we concentrate on the CP-even Higgs sector in the following.

The mass eigenstates $h_a$ are related to the gauge eigenstates using
the unitary matrix $S_{aA}$ as follows:
\begin{eqnarray}
  \begin{pmatrix}
    h_1 \\ 
    h_2 \\ 
    h_3
  \end{pmatrix}
  &=&
  \begin{pmatrix}
    S_{1d} & S_{1u} & S_{1s} \\ 
    S_{2d} & S_{2u} & S_{2s} \\ 
    S_{3d} & S_{3u} & S_{3s}
  \end{pmatrix}
  \begin{pmatrix}
    h^0_d \\ 
    h^0_u \\ 
    s 
  \end{pmatrix},
  \label{fig:mixing_shH}
\end{eqnarray}
where $h^0_d$, $h^0_u$, and $s$ are the real scalars in the gauge
eigenstates $H_d$, $H_u$, and $S$, respectively.  In the following, we
consider the case that $h_2$ is (almost) the singlet field, so that
$|S_{2s}|\gg|S_{1s}|$, $|S_{3s}|$, $|S_{2d}|$, $|S_{2u}|$.

Now, we are at the position to discuss production and decay processes
of the singlet Higgs at the LHC.  The gluon fusion cross section for
the process $pp\rightarrow h_2\rightarrow F$ (with $F$ denoting the
final state of the decay of $h_2$) is given by
\begin{eqnarray}
  \sigma(pp\rightarrow h_2\rightarrow F) = 
  \frac{\Gamma(h_2\to F)}{\Gamma_{\rm total}}
  \sigma(pp\rightarrow h_2),
\end{eqnarray}
with
\begin{eqnarray}
  \sigma(pp\rightarrow h_2) = 
  \Gamma(h_2 \to g g) 
  \frac{\pi^2}{8 m_{h_2} s} 
  \int_0^1 dx_1 \int_0^1 dx_2 \delta(x_1x_2 - m_{h_2}^2/s)
  g(x_1) g(x_2),
  \label{eq:sigma_pph2F}
\end{eqnarray}
where $\Gamma(h_2\to F)$, $\Gamma(h_2\to gg)$ and $\Gamma_{\rm total}$
are the partial decay widths of the decay processes $h_2 \to F$, $h_2
\to gg$ and the total decay width of $h_2$, respectively. In
Eq.(\ref{eq:sigma_pph2F}), $\sqrt{s}$ is the center
of mass energy and $g(x)$ is the gluon distribution in
the proton. In the present set up, the process $h_2\rightarrow F$ may
be induced by the mixing with $H_d$ and $H_u$ as well as by the loop
diagrams with $D_i$ and $\bar{D}_i$ inside the loop.

With a relevant choice of parameters, we found that the process
$h_2\rightarrow gg$ is dominated by the loop diagrams with $D_i$ and
$\bar{D}_i$ inside the loop, while the decay processes into the weak
boson pairs $h_2\rightarrow VV$ (with $VV=ZZ$ or $W^+W^-$) are due to
the mixing effect.  Thus,
\begin{eqnarray}
  \Gamma(h_2\rightarrow gg) \simeq |S_{2s}|^2 
  \frac{m_{h_2}^3 \alpha_s^2}{144 \pi^3 v_s^2} 
  \left| N_{\bf 5} I_D \right|^2,
  \label{eq:gammahgg}
\end{eqnarray}
and
\begin{eqnarray}
  \Gamma(h_2\rightarrow VV) \simeq 
  |S_{2d}\cos\beta+S_{2u}\sin\beta|^2 \Gamma(h_{\rm SM}\rightarrow VV),
\end{eqnarray}
where $\Gamma(h_{\rm SM}\rightarrow VV)$ is the partial decay width of
the SM Higgs boson (with the mass of $m_{h_2}$).  In
Eq.(\ref{eq:gammahgg}), 
\begin{eqnarray}
  I_D= 3 \left[ 2 x_D + x_D(4 x_D - 1) f(x_D) \right],
\end{eqnarray}
where 
\begin{eqnarray}
  f(x) =
  \left\{ \begin{array}{ll}
      - 2 \arcsin^2 (1/2\sqrt{x}) & 
      {\rm for} \quad x > 1/4
      \\ \\
      \displaystyle{
        \frac{1}{2}  \left[
          \ln \left( \frac{ 1 + \sqrt{1-4x} }
            { 1 - \sqrt{1-4x} } \right)
          - i \pi \right]^2
      }
      & 
      {\rm for} \quad x < 1/4
    \end{array} \right. ,
\end{eqnarray}
and $x_D= m_D^2/m_{h_2}^2 = y_D^2 v_s^2/m_{h_2}^2$ \cite{Barger:1987nn}.  In fact,
$I_D\simeq 1$ when $x_D\gtrsim 1$, so we approximate $I_D=1$ in our
numerical calculation.  Then the results are insensitive to the choice
of $y_D$ (as far as $m_D$ is assumed to be larger than $m_{h_2}$).

One important point is that the decay width $\Gamma(h_2\rightarrow
gg)$ (and hence the gluon fusion cross section) is proportional to
$N_{\bf 5}^2$; this is because the amplitude is proportional to the
number of particles inside the loop.  Thus, as we increase the number
of vector-like chiral multiplets, the signal of singlet
Higgs production process is enhanced.

Using the formulae given above as well as the numerical package
NMSSMtools \cite{NMSSMtools}, we calculate the cross sections for the Higgs
production processes in the present set up. We have calculated the
decay width of the SM Higgs to gluon-gluon mode at the leading order.
Decay widths of the SM Higgs (except for the $gg$ mode) are calculated by
HDECAY package \cite{Djouadi:1997yw}.

First, we show the cross section for the process $pp\rightarrow
h_2\rightarrow VV$ normalized by the corresponding SM process:
\begin{eqnarray}
  R_{VV} \equiv
  \frac{\sigma(pp\rightarrow h_2\rightarrow VV)}
  {\sigma(pp\rightarrow h_{\rm SM}\rightarrow VV)}.
\end{eqnarray}
In our analysys, only the gluon fusion process, which is the dominant Higgs production process 
at the LHC, is taken into account. 

%--------------------------------------------------------->>>FIG
\begin{figure}
  \begin{center}
    \includegraphics[origin=b, angle=0,width=8.5cm]{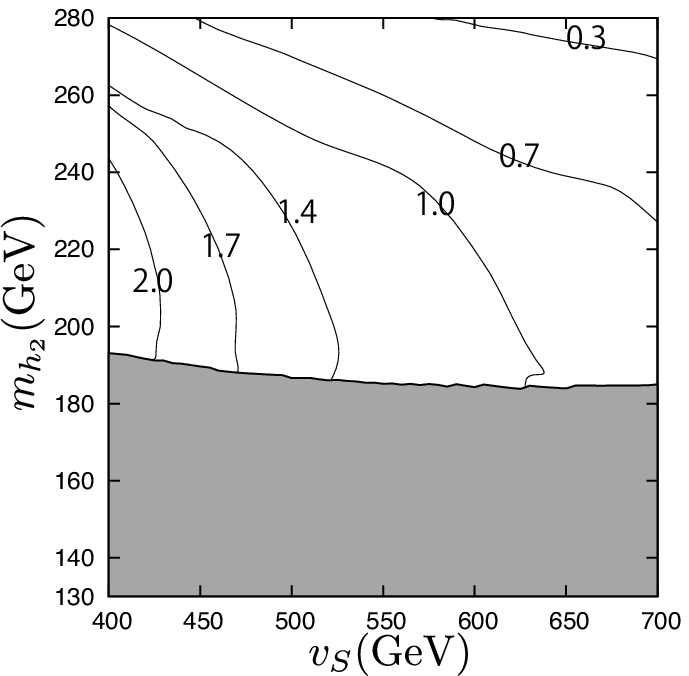}
    \caption{The contours of constant $R_{ZZ}$ on $v_s$ vs.\ $m_{h_2}$
      plane for $\sqrt{s} = 7 {\rm TeV}$.  Here, we take $N_{\bf 5}=4$, $\tan\beta=3$ and $\mu_{\rm
        eff}=250\ {\rm GeV}$. In the shaded region, the lightest Higgs boson 
      mass is less than $114.4$ GeV }
    \label{fig:Rzz_tb3}
\vspace{1cm}
    \includegraphics[origin=b, angle=0,width=8.5cm]{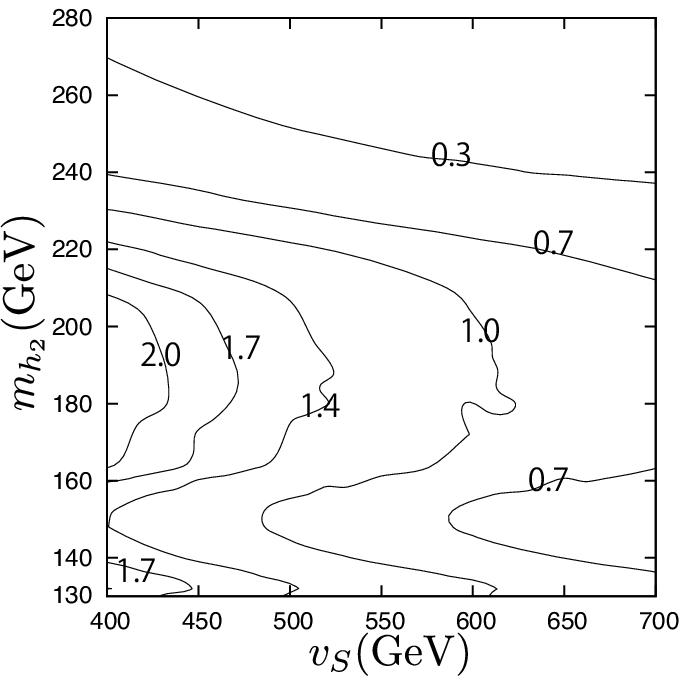}
    \caption{The contours of constant $R_{ZZ}$ on $v_s$ vs.\ $m_{h_2}$
      plane for $\sqrt{s} = 7 {\rm TeV}$.  Here, we take $N_{\bf 5}=4$, $\tan\beta=5$ and $\mu_{\rm
        eff}=110\ {\rm GeV}$.}
    \label{fig:Rzz_tb5}
  \end{center}
\end{figure}
%---------------------------------------------------------<<<FIG

In Figs.\ \ref{fig:Rzz_tb3} and \ref{fig:Rzz_tb5}, we show the
contours of constant $R_{ZZ}$ for $\sqrt{s} = 7 {\rm TeV}$; we have checked that the behavior of
$R_{W^+W^-}$ is almost identical.  In the calculation, we have taken
$\tan\beta=3$ and $\mu_{\rm eff} = 250\ {\rm GeV}$ (Fig.\
\ref{fig:Rzz_tb3}) and $\tan\beta=5$ and $\mu_{\rm eff} = 110\ {\rm
  GeV}$ (Fig.\ \ref{fig:Rzz_tb5}).  In Fig.\ \ref{fig:Rzz_tb3}, the
region with $m_{h_2}\lesssim 195\ {\rm GeV}$ is excluded because the
lightest Higgs becomes lighter than the LEP bound (i.e., $114.4\ {\rm
  GeV}$) \cite{Barate:2003sz}.  Such a result is partly because the choice
of $\tan\beta$ parameter.  Indeed, if we adopt a slightly larger value
of $\tan\beta$, the region with $m_{h_2}\lesssim 195\ {\rm GeV}$
becomes allowed.  (See Fig. \ref{fig:Rzz_tb5}.)  In the case of Fig.
\ref{fig:Rzz_tb5}, $R_{VV}$ is suppressed for $m_{h_2}\gtrsim 220\
{\rm GeV}$; in such a region, the decay mode of $h_2$ into
Higgsino-like fermions becomes kinematically open.

In both figures, one can see that $R_{VV}\gtrsim 1$ can be realized in
large fraction of the parameter space.  Thus, the singlet production
process may mimic the SM-Higgs production at the LHC.  Notice that the
masses of colored fermions are given by $m_D=y_Dv_s$ and hence,
assuming the perturbativity of the Yukawa coupling constant, the
colored fermions are expected to be lighter than $\sim v_s$.  Thus, for our
scenario to work, there should exist colored fermions with masses of
$O(100\ {\rm GeV})$.  The detailed upper bound on the colored fermion
masses depends on various parameters; for the case of Fig.\
\ref{fig:Rzz_tb3}, for example, $R_{VV}=1$ requires $m_D\lesssim 630\
{\rm GeV}$ and $500\ {\rm GeV}$ for $m_{h_2}=200\ {\rm GeV}$ and
$m_{h_2}=250\ {\rm GeV}$, respectively.  Thus the search for the
colored fermions should give a crucial test of the present scenario.
The experimental signal of the production of such colored particles
depends how they decay.  In the present set up, $D$ and $\bar{D}$
may slightly mix with down-type quarks via Yukawa-type interactions.
In such a case, they decay into an up-type light quark and $W^\pm$-boson.  The
current lower bound on the masses of such particles is about $400\
{\rm GeV}$ for $N_{\bf 5}=4$ \cite{q4_bound}.

\begin{figure}
  \begin{center}
    \includegraphics[origin=b, angle=0,width=8.5cm]{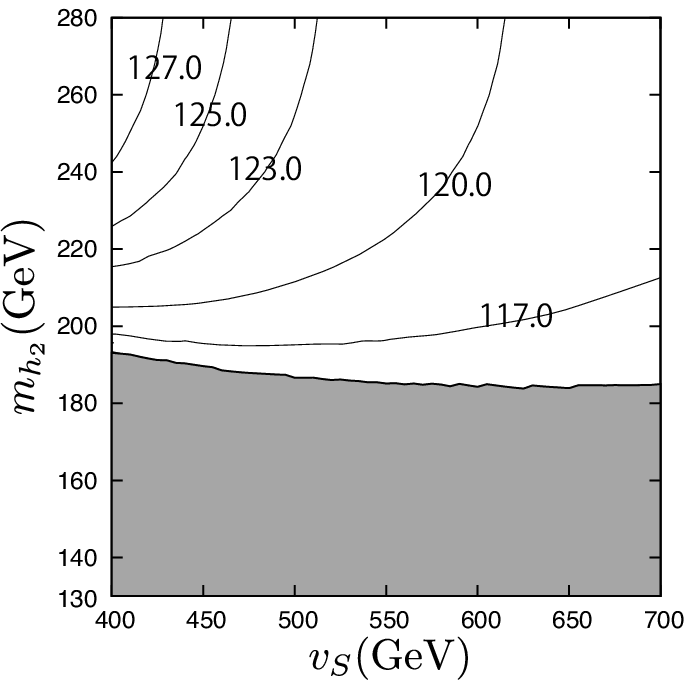}
    \caption{The contours of constant lightest Higgs mass on $v_s$ vs.\ $m_{h_2}$
      plane.  Here, we take $N_{\bf 5}=4$, $\tan\beta=3$ and $\mu_{\rm
        eff}=250\ {\rm GeV}$. In the shaded region, the lightest Higgs boson 
      mass is less than $114.4$ GeV.}
    \label{fig:mh_tb3}
\vspace{1cm}
    \includegraphics[origin=b, angle=0,width=8.5cm]{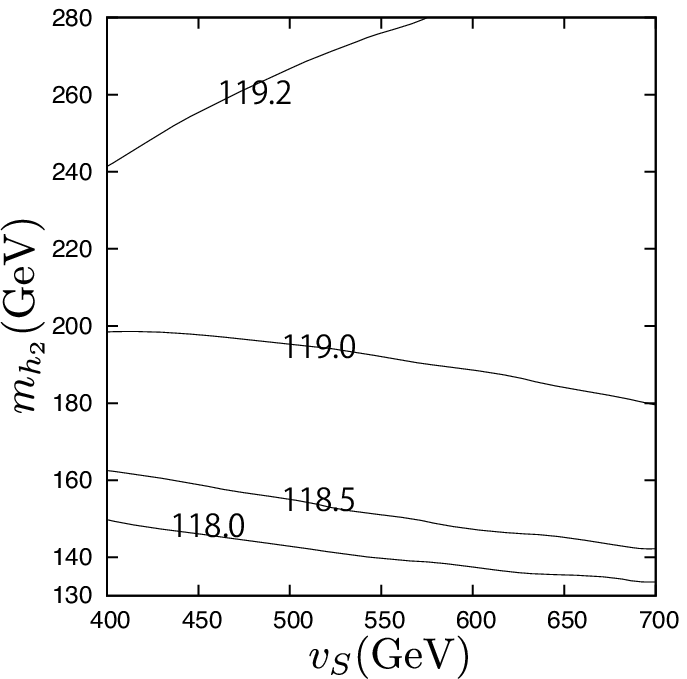}
    \caption{The contours of constant lightest Higgs mass on $v_s$ vs.\ $m_{h_2}$
      plane.  Here, we take $N_{\bf 5}=4$, $\tan\beta=5$ and $\mu_{\rm
        eff}=110\ {\rm GeV}$.}
    \label{fig:mh_tb5}
  \end{center}
\end{figure}

Because the lightest Higgs mass is also an important parameter, we
have calculated $m_{h_1}$ in both cases.  The results are shown in
Figs.\ \ref{fig:mh_tb3} and \ref{fig:mh_tb5}.
%--------------------------------------------------------->>>FIG
%---------------------------------------------------------<<<FIG

Finally, let us comment on the decay mode $h_2\to\gamma\gamma$.  As in
the case of the process $h_2\to gg$, the decay width for the process
$h_2\to\gamma\gamma$ may be also affected by the loop diagrams with
the vector-like fermions inside the loop.  Importantly, for
$h_2\to\gamma\gamma$, the $W^\pm$-loop diagram contribution is
numerically significant as in the case of SM Higgs boson.  Even though
such a contribution is suppressed by the mixing factor
($S_{2d}\cos\beta+S_{2u}\sin\beta$), we have checked that the effect of
$W^\pm$-loop diagram is comparable to that of vector-like-fermion loop
in the parameter region we have studied.  Thus, the branching ratio
$Br(h_2\rightarrow\gamma\gamma)$ may significantly deviate from that
of the SM Higgs boson.

\noindent
{\it Acknowledgments}:
This work is supported by Grant-in-Aid for Scientific research from
the Ministry of Education, Science, Sports, and Culture (MEXT), Japan,
No.\ 22540263 (T.M.), No.\ 22244021 (M.A. and T.M.) and No.\ 22-7585(N.Y.).

%%%%%%%%%%%%%%%%%%%%%%%%%%%%%%%%%%%%%%%%%%%%%%%%%%%%%%%%%%%%%%%%%%%%%%%%%%%%%%%%%

\end{document}